\documentclass[aps,prl,preprint,tightenlines,superscriptaddress,showpacs,byrevtex,twocolumn]{revtex4}
\usepackage{graphicx}
\usepackage{dcolumn}
\usepackage{bm}
\usepackage{amsmath}
\usepackage{amssymb}
\usepackage{color}

\begin{document}

\preprint{\vbox{ 
								 \hbox{BELLE-CONF-0769}
                 \hbox{BELLE Preprint 2008-10}
                 \hbox{KEK Preprint 2008-4}
%                 \hbox{Draft version PRD(RC) v1.0}
                              }}

\title{\quad\\
[0.5cm] Search for $B \to \pi \ell^+\ell^-$ Decays at Belle}

%%% Paper:    B -> pi l l
%%% Journal:  Physical Review D (Rapid Comm) %%% Contacts: J.-T. Wei (alan1215@hep1.phys.ntu.edu.tw)
%%%           K.-F. Chen (kfjack@hep1.phys.ntu.edu.tw)
%%% Non-responding authors or those who said NO are commented out.
%%% ====================================================================
%%% Click the RELOAD button on your web browser to see the updated file.
%%% ====================================================================
%%% Use \input{author} to insert this material into your latex file.
%%%%% Force institutions to appear in alphabetical order when typeset.
\affiliation{Budker Institute of Nuclear Physics, Novosibirsk} %%%\affiliation{Chiba University, Chiba} \affiliation{University of Cincinnati, Cincinnati, Ohio 45221} %%%\affiliation{Department of Physics, Fu Jen Catholic University, Taipei} %%%\affiliation{Justus-Liebig-Universit\"at Gie\ss{}en, Gie\ss{}en} \affiliation{The Graduate University for Advanced Studies, Hayama} %%%\affiliation{Gyeongsang National University, Chinju} \affiliation{Hanyang University, Seoul} \affiliation{University of Hawaii, Honolulu, Hawaii 96822} \affiliation{High Energy Accelerator Research Organization (KEK), Tsukuba} %%%\affiliation{Hiroshima Institute of Technology, Hiroshima} %%%\affiliation{University of Illinois at Urbana-Champaign, Urbana, Illinois 61801} \affiliation{Institute of High Energy Physics, Chinese Academy of Sciences, Beijing} \affiliation{Institute of High Energy Physics, Vienna} \affiliation{Institute of High Energy Physics, Protvino} \affiliation{Institute for Theoretical and Experimental Physics, Moscow} \affiliation{J. Stefan Institute, Ljubljana} \affiliation{Kanagawa University, Yokohama} \affiliation{Korea University, Seoul} %%%\affiliation{Kyoto University, Kyoto} \affiliation{Kyungpook National University, Taegu} \affiliation{\'Ecole Polytechnique F\'ed\'erale de Lausanne (EPFL), Lausanne} \affiliation{Faculty of Mathematics and Physics, University of Ljubljana, Ljubljana} \affiliation{University of Maribor, Maribor} \affiliation{University of Melbourne, School of Physics, Victoria 3010} \affiliation{Nagoya University, Nagoya} %%%\affiliation{Nara Women's University, Nara} \affiliation{National Central University, Chung-li} \affiliation{National United University, Miao Li} \affiliation{Department of Physics, National Taiwan University, Taipei} \affiliation{H. Niewodniczanski Institute of Nuclear Physics, Krakow} \affiliation{Nippon Dental University, Niigata} \affiliation{Niigata University, Niigata} \affiliation{University of Nova Gorica, Nova Gorica} \affiliation{Osaka City University, Osaka} \affiliation{Osaka University, Osaka} \affiliation{Panjab University, Chandigarh} %%%\affiliation{Peking University, Beijing} %%%\affiliation{University of Pittsburgh, Pittsburgh, Pennsylvania 15260} %%%\affiliation{Princeton University, Princeton, New Jersey 08544} %%%\affiliation{RIKEN BNL Research Center, Upton, New York 11973} %%%\affiliation{Saga University, Saga} \affiliation{University of Science and Technology of China, Hefei} \affiliation{Seoul National University, Seoul} %%%\affiliation{Shinshu University, Nagano} \affiliation{Sungkyunkwan University, Suwon} \affiliation{University of Sydney, Sydney, New South Wales} %%%\affiliation{Tata Institute of Fundamental Research, Mumbai} \affiliation{Toho University, Funabashi} \affiliation{Tohoku Gakuin University, Tagajo} %%%\affiliation{Tohoku University, Sendai} \affiliation{Department of Physics, University of Tokyo, Tokyo} \affiliation{Tokyo Institute of Technology, Tokyo} \affiliation{Tokyo Metropolitan University, Tokyo} \affiliation{Tokyo University of Agriculture and Technology, Tokyo} %%%\affiliation{Toyama National College of Maritime Technology, Toyama} \affiliation{Virginia Polytechnic Institute and State University, Blacksburg, Virginia 24061} \affiliation{Yonsei University, Seoul} % \author{K.~Abe}\affiliation{High Energy Accelerator Research Organization (KEK), Tsukuba} % KEK
   \author{J.-T.~Wei}\affiliation{Department of Physics, National Taiwan University, Taipei} % Taiwan
   \author{K.-F.~Chen}\affiliation{Department of Physics, National Taiwan University, Taipei} % Taiwan
   \author{I.~Adachi}\affiliation{High Energy Accelerator Research Organization (KEK), Tsukuba} % KEK
   \author{H.~Aihara}\affiliation{Department of Physics, University of Tokyo, Tokyo} % Tokyo % \author{D.~Anipko}\affiliation{Budker Institute of Nuclear Physics, Novosibirsk} % BINP
   \author{K.~Arinstein}\affiliation{Budker Institute of Nuclear Physics, Novosibirsk} % BINP % \author{T.~Aso}\affiliation{Toyama National College of Maritime Technology, Toyama} % Toyama
   \author{V.~Aulchenko}\affiliation{Budker Institute of Nuclear Physics, Novosibirsk} % BINP
   \author{T.~Aushev}\affiliation{\'Ecole Polytechnique F\'ed\'erale de Lausanne (EPFL), Lausanne}\affiliation{Institute for Theoretical and Experimental Physics, Moscow} % ITEP % \author{T.~Aziz}\affiliation{Tata Institute of Fundamental Research, Mumbai} % Tata % \author{S.~Bahinipati}\affiliation{University of Cincinnati, Cincinnati, Ohio 45221} % Cincinnati
   \author{A.~M.~Bakich}\affiliation{University of Sydney, Sydney, New South Wales} % Sydney
   \author{V.~Balagura}\affiliation{Institute for Theoretical and Experimental Physics, Moscow} % ITEP % \author{Y.~Ban}\affiliation{Peking University, Beijing} % Peking % \author{S.~Banerjee}\affiliation{Tata Institute of Fundamental Research, Mumbai} % Tata % \author{E.~Barberio}\affiliation{University of Melbourne, School of Physics, Victoria 3010} % Melbourne % \author{M.~Barbero}\affiliation{University of Hawaii, Honolulu, Hawaii 96822} % Hawaii % \author{A.~Bay}\affiliation{\'Ecole Polytechnique F\'ed\'erale de Lausanne (EPFL), Lausanne} % Lausanne % \author{I.~Bedny}\affiliation{Budker Institute of Nuclear Physics, Novosibirsk} % BINP
   \author{K.~Belous}\affiliation{Institute of High Energy Physics, Protvino} % Protvino % \author{V.~Bhardwaj}\affiliation{Panjab University, Chandigarh} % Panjab
% ...
% BINP \collaboration{The Belle Collaboration}

\begin{abstract}
We present a search for the decays $B^+ \to \pi^+ \ell^+\ell^-$ and $B^0 \to \pi^0 \ell^+\ell^-$, 
where $\ell^+\ell^-$ is either a $\mu^+ \mu^-$ or $e^+e^-$ pair,
with a data sample of 657 million $B\overline{B}$ pairs 
collected with the Belle detector at the KEKB $e^+e^-$ collider. 
Signal events are reconstructed from a charged or a neutral pion candidate and a pair of oppositely charged electrons or muons.
%By performing maximum likelihood fits to the reconstructed candidates, 
No significant signal is observed and 
%we set upper limits on the branching fractions
we set an upper limit on the isospin-averaged branching fraction
$\mathcal{B}(B \to \pi\ell^+\ell^-) < 6.2 \times 10^{-8}$
at the 90\% confidence level. 

\end{abstract}
\pacs{13.25 Hw, 13.20 He}

\collaboration{Belle Collaboration} \noaffiliation

\maketitle
    
\tighten    

\begin{figure}[htpb]
\begin{center}
\includegraphics[width=5.5cm]{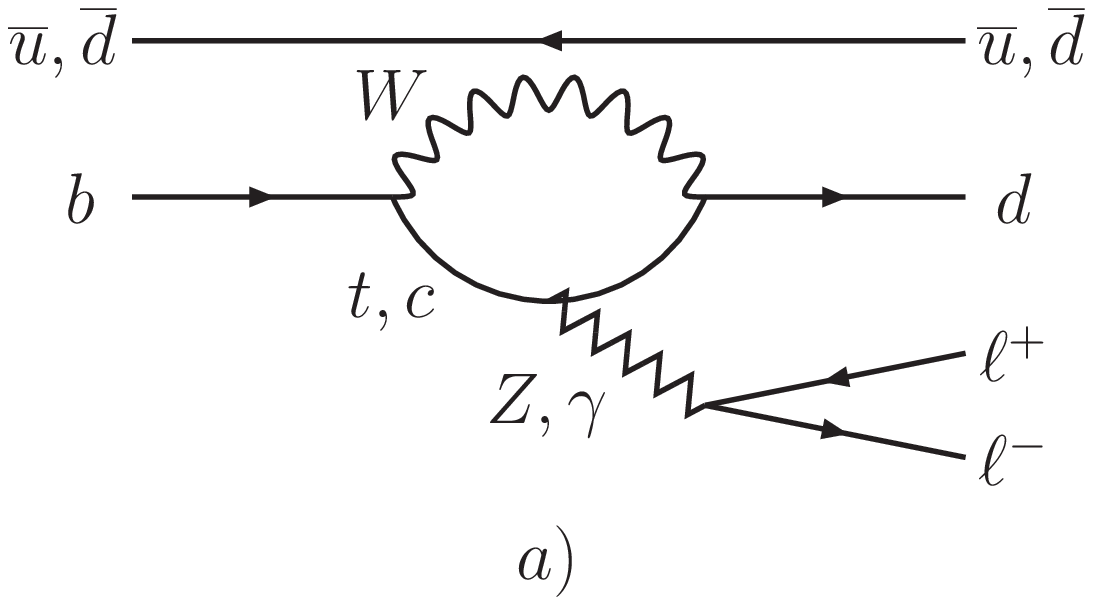} 
%\hskip -0.2cm
\includegraphics[width=5.5cm]{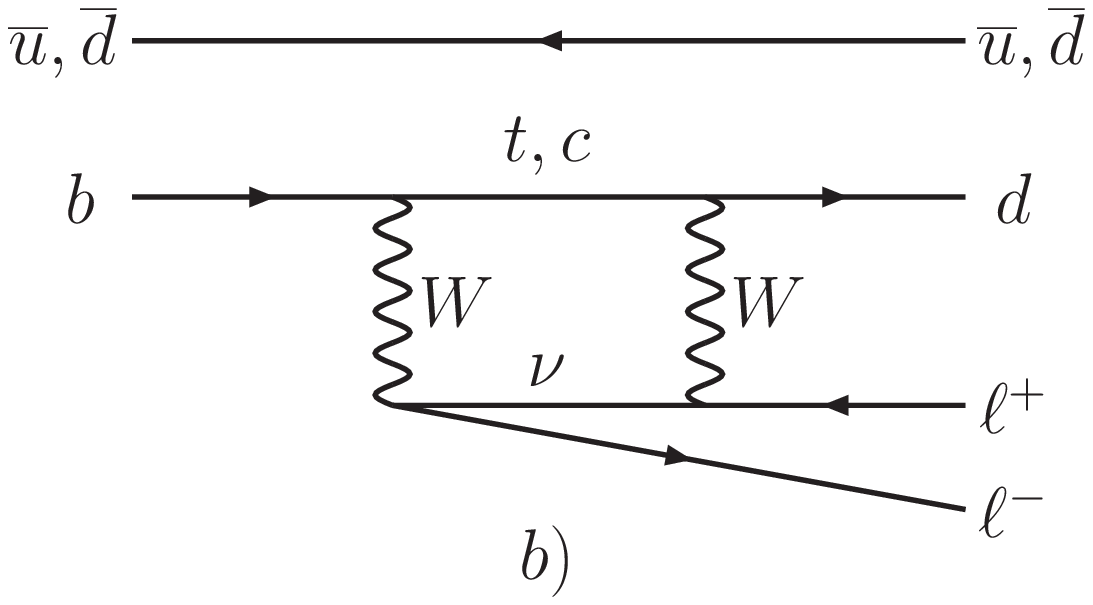}
\end{center}
\caption{The quark-level diagrams for $B \to \pi \ell^+\ell^-$ decays.}
\label{fig:diagrams}
\end{figure}

In the Standard Model (SM), the decays $B \to \pi \ell^+\ell^-$ are forbidden at tree level and 
only proceed through flavor-changing neutral currents (FCNC). 
The possible lowest-order $b \to d \ell^+\ell^-$ processes are either a $Z$/$\gamma$ penguin or 
a box diagram, as shown in Fig.~\ref{fig:diagrams}. 
FCNC $b \to (s,d) \ell^+\ell^-$ decays can provide 
stringent tests of the SM in the flavor physics sector.
For example, in the future it may be possible to measure 
%especially for the possibility of measuring 
the dilepton invariant mass spectrum and the forward-backward asymmetry in the 
center-of-mass system of the lepton pair, 
which can provide information on the coefficients of new operators 
associated with theoretical models~\cite{ref:ali}.
Most of the experimental studies and theoretical predictions are focused on $b \to s \ell^+\ell^-$ decays~\cite{ref:kll_exp,ref:kll_th}; 
however, signatures due to new physics may be observed in
$b \to d \ell^+\ell^-$ decays even if no such signature is found in $b \to s \ell^+\ell^-$.
%However, if there are some New Physics effects observed in the interactions, the same signature is expected in $b \to d \ell^+\ell^-$ 
%decays as an independent attestation.

%By taking the $b \to s \ell^+\ell^-$ process as a reference, a much smaller branching fraction for 
%$b \to d \ell^+\ell^-$ is expected, with a suppression factor of
The process $b \to d \ell^+\ell^-$ is suppressed by a factor of
$|V_{td}/V_{ts}|^2 \approx 0.04$ relative to $b \to s \ell^+\ell^-$,
where $V_{td}$ and $V_{ts}$ are the elements of 
the Cabibbo-Kobayashi-Maskawa quark-mixing matrix~\cite{ref:pdg}.
%A prediction in the framework of two Higgs doublet model shows a branching fraction of 
%A prediction in the framework of SM shows a branching fraction of
%$3.3 \times 10^{-8}$ for the exclusive $B^+ \to \pi^+ \ell^+\ell^-$ decay, 
The corresponding branching fraction of the exclusive 
$B^+ \to \pi^+ \ell^+\ell^-$ mode
is expected to be $3.3 \times 10^{-8}$, 
which agrees with the simple power counting expectation~\cite{ref:Aliev1999}. 
%Assumption of the isospin symmetry in the $B^0 \to \pi^0 \ell^+\ell^-$ decays 
%causes the branching fractions to decrease by a factor of 
Assuming isospin symmetry, the branching fraction for $B^0 \to \pi^0 \ell^+\ell^-$ 
is smaller by another factor of 
$1/2 \times \tau_{B^0}/\tau_{B^+}$ for these decays, 
where $\tau_{B^0}$ and $\tau_{B^+}$ are the neutral and charged $B$-meson lifetimes, respectively.
%With an assumption of isospin symmetry, the branching fraction for 
%the $B^0 \to \pi^0 \ell^+\ell^-$ process is smaller by a factor of 
%$1/2 \times \tau_{B^0}/\tau_{B^+}$,
%where $\tau_{B^0}$ and $\tau_{B^+}$ are the charged and neutral $B$-meson lifetimes, respectively. 
Thus for SM branching fractions, only a few signal candidates are expected 
in our data sample.
%Due to the extremely small branching fractions for these decays, previous experiemental result from BaBar~\cite{Aubert:2006ah} 
%shows no hint of signal. 

In this paper, we perform a search for the $B^+ \to \pi^+ \ell^+\ell^-$ and $B^0 \to \pi^0 \ell^+\ell^-$ decays, 
where $\ell^+\ell^-$ stands for a $\mu^+ \mu^-$ or $e^+e^-$ pair. A data sample of 657 million $B\overline{B}$ pairs 
collected with the Belle detector at the KEKB $e^+e^-$ collider is examined. 
Charge-conjugate decays are implied throughout this paper.

The Belle detector is a large-solid-angle magnetic spectrometer located at
the KEKB collider~\cite{ref:KEKB}, and consists of a silicon vertex detector
(SVD), a 50-layer central drift chamber (CDC), an array of 
aerogel threshold Cherenkov counters (ACC), a barrel-like arrangement of
time-of-flight scintillation counters (TOF), and an electromagnetic
calorimeter comprised of CsI(Tl) crystals (ECL) located inside a
superconducting solenoid that provides a 1.5~T magnetic field. An iron
flux-return located outside the coil is instrumented to detect $K_L^0$
mesons and to identify muons (KLM). The detector is described in detail
elsewhere~\cite{ref:belle_detector}.

%All charged tracks are required to have a 
We consider all tracks that have a
maximum distance to the interaction point (IP) of 5 cm in the beam direction ($z$) and
of 0.5 cm in the transverse plane ($r$--$\phi $).
We select $\pi^\pm$ candidates from charged tracks
having a kaon likelihood ratio less than 0.4;
the kaon likelihood ratio is defined by 
$\mathcal{R}_{K}\equiv \mathcal{L}_{K}/(\mathcal{L}_{K}+\mathcal{L}_{\pi})$,
where $\mathcal{L}_{K}$ ($\mathcal{L}_{\pi }$) denotes a likelihood
that combines measurements from the ACC, the TOF, 
and $dE/dx$ from the CDC for the $K^{\pm}$ ($\pi^{\pm}$) tracks.
The selection efficiency is about 89\% for pions while it removes 91\% of kaons.
In addition to the information included in the kaon likelihood ratio, 
muon (electron) candidates are
required to be associated with KLM detector hits (ECL calorimeter showers).
We define the likelihood ratio $\mathcal{R}_{x}$ ($x$ denotes $\mu$ or $e$) as
%$\mathcal{R}_{x}\equiv \mathcal{L}_{x}/({\mathcal{L}_{x}}+{\mathcal{L}_{K}}+{\mathcal{L}_{\pi}})$,
%where $\mathcal{L}_{x}$, $\mathcal{L}_{K}$ and $\mathcal{L}_{\pi}$ 
$\mathcal{R}_{x}\equiv \mathcal{L}_{x}/({\mathcal{L}_{x}}+{\mathcal{L}_{not-x}})$,
where $\mathcal{L}_{x}$ and $\mathcal{L}_{not-x}$ 
are the likelihood measurements from the relevant detectors~\cite{ref:lid}.
We select $\mu^\pm$ candidates with 
%a minimum muon likelihood ratio of 0.9
$\mathcal{R}_{\mu} >$ 0.9
if the momentum of $\mu^\pm$ is greater than 1 GeV/$c$;
for the $\mu^\pm$ candidates with lower momentum (0.7--1.0 GeV/$c$),
%the muon likelihood ratio 
$\mathcal{R}_{\mu}$ is required to be greater than 0.97.
These requirements retain about 80\% of muons while removing 98.5\% of pions.
Candidates for $e^{\pm}$ are required to have a minimum momentum of 0.4 GeV/$c$,  
%a minimum electron likelihood ratio of 0.8; 
$\mathcal{R}_{e} >$ 0.9, and $\mathcal{R}_{\mu} <$ 0.8.
%and the ones with $\mathcal{R}_{\mu} >$ 0.8 are excluded from the study.
These requirements keep about 90\% of electrons while they remove 99.7\% of pions.
Bremsstrahlung photons emitted by the electrons are recovered by adding 
neutral clusters found within a 50 mrad cone along the electron direction. 
The energy of the additional photon is required to be less than 0.5 GeV.
For $\pi^{0} \to \gamma\gamma$ candidates, a minimum photon energy of 50 MeV
is required and the invariant mass must be in the range 115 MeV/$c^2$ $< M(\gamma\gamma) < $ 152 MeV/$c^2$, corresponding to $\pm$ 3 $\sigma$ for the $\pi^0$ reconstruction.
%%pi0 spectrum from MC = $134.6 \pm 11.42$ MeV
Requirements on the photon energy asymmetry, $|E_{\gamma}^1 - E_{\gamma}^2|/(E_{\gamma}^1 + E_{\gamma}^2)<0.9$,
and the minimum momentum of the $\pi^0$ candidate, $p_{\pi^0} >$ 200 MeV/$c$,
are introduced to suppress the combinatorial background. 

$B$-meson candidates are reconstructed in one of the following modes: 
$B^+ \to \pi^+ e^+ e^-$, $B^+ \to \pi^+ \mu^+ \mu^-$, $B^0 \to \pi^0 e^+ e^-$, or $B^0 \to \pi^0 \mu^+ \mu^-$.
Signal candidates are selected using the
beam-energy constrained mass $M_{\mathrm{bc}} \equiv \sqrt{E_{\mathrm{beam}}^{2} - p_{B}^{2}}$ 
and the energy difference $\Delta E \equiv E_{B} - E_{\mathrm{beam}}$, 
where $E_{B}$ and $p_{B}$ are the reconstructed energy and momentum
of the $B$ candidate in the $\Upsilon(4S)$ center-of-mass
(CM) frame, and $E_{\mathrm{beam}}$ is the beam energy in this frame.
We require $B$-meson candidates to be within the region
$M_{\mathrm{bc}}>5.20$~GeV/$c^2$ and $-0.1$~GeV~$<\Delta E< 0.3$~GeV 
($-0.15$~GeV~$<\Delta E< 0.3$~GeV) for the
$B^+ \to \pi^+ \ell^+ \ell^-$ ($B^0 \to \pi^0 \ell^+ \ell^-$) decays.
The signal region is defined by 5.27~GeV/$c^2$ $< M_{\mathrm{bc}} <$ 5.29~GeV/$c^2$, 
$-0.035$ ($-0.08$) GeV $< \Delta E <$ 0.035 GeV for the $\pi^+\mu^+\mu^-$ ($\pi^0\mu^+\mu^-$) mode,
and $-0.055$ ($-0.1$) GeV $< \Delta E <$ 0.035 GeV for the $\pi^+e^+e^-$ ($\pi^0e^+e^-$) mode.
%\textcolor{blue}{
The lower $\Delta E$ bound for the candidate region is designed to 
exclude possible contamination from similar $B$ decays, e.g. $B\to K^{*} \ell \ell$,
while the wider region in $\Delta E$ for $B^0 \to \pi^0 \ell^+ \ell^-$ decays 
is chosen to include the long tail of the signal.%}

%The dominant background is continuum
%$e^+e^- \to q\overline{q}$ production ($q=u,d,c,s$).
The dominant source of background is continuum $e^+e^- \to q\overline{q}$ events ($q=u,d,c,s$).
A Fisher discriminant including 16 modified Fox-Wolfram moments~\cite{ref:sfw} 
is used to exploit the differences between
the event shapes for continuum $q\overline{q}$ production
(jet-like) and for $B$ decay (spherical) in the $e^+e^-$ rest frame.
We combine 1) the Fisher discriminant,
2) the angle between the momentum vector of reconstructed $B$ candidate and beam direction ($\cos\theta_B$), 
and 3) the distance in the $z$ direction between the candidate $B$ vertex and a vertex position 
formed by the charged tracks that are not associated with the candidate $B$-meson 
into a single likelihood ratio
$\mathcal{R} = {\cal L}_s/({\cal L}_s + {\cal L}_{q\overline{q}})$, where
${\cal L}_s$ (${\cal L}_{q\overline{q}}$) denotes the signal (continuum) likelihood.
The background from $B \to \ell \nu X$ decays is also found to be large. An additional 
likelihood ratio $\mathcal{R}_{B} = {\cal L}_s/({\cal L}_s + {\cal L}_{B\overline{B}})$ 
including $\cos\theta_B$ and the overall missing energy is introduced, where 
${\cal L}_{B\overline{B}}$ is the likelihood for $B\overline{B}$ events.

Continuum background suppression is improved
by including $B$-flavor tagging information~\cite{Kakuno:2004cf}, 
which is parameterized by a discrete variable $q_{\rm tag}$ indicating 
the flavor of the tagging $B$-meson candidate and a quality parameter $r$ 
(ranging from 0 for no flavor information to 1 for unambiguous flavor assignment).
Selection criteria for $\mathcal{R}$ and $\mathcal{R}_{B}$ are determined by maximizing
the value of $S/\sqrt{S+B}$, where $S$ and $B$ denote the expected yields 
of signal and background events in the signal region, respectively,
in different $q_{\rm rec} \cdot q_{\rm tag} \cdot r$ regions,
where $q_{\rm rec}$ is the charge of reconstructed $B$ candidate.
%Events with a value close to $-1$ ($+1$) of $q_{\rm rec} \cdot q_{\rm tag} \cdot r$ ($r$) 
%Events with a value close to $\pm 1$ of $q_{\rm tag} \cdot r$ 
Events with $q_{\rm rec} \cdot q_{\rm tag} \cdot r$ close to $-1$
are considered to be well-tagged and are unlikely to be from continuum processes.
For neutral $B$-meson decays, only the dependence on $r$ is considered. 

\begin{figure}[htpb]
\begin{center}
\vskip -0.5cm
\includegraphics[width=5.5cm,height=5.5cm]{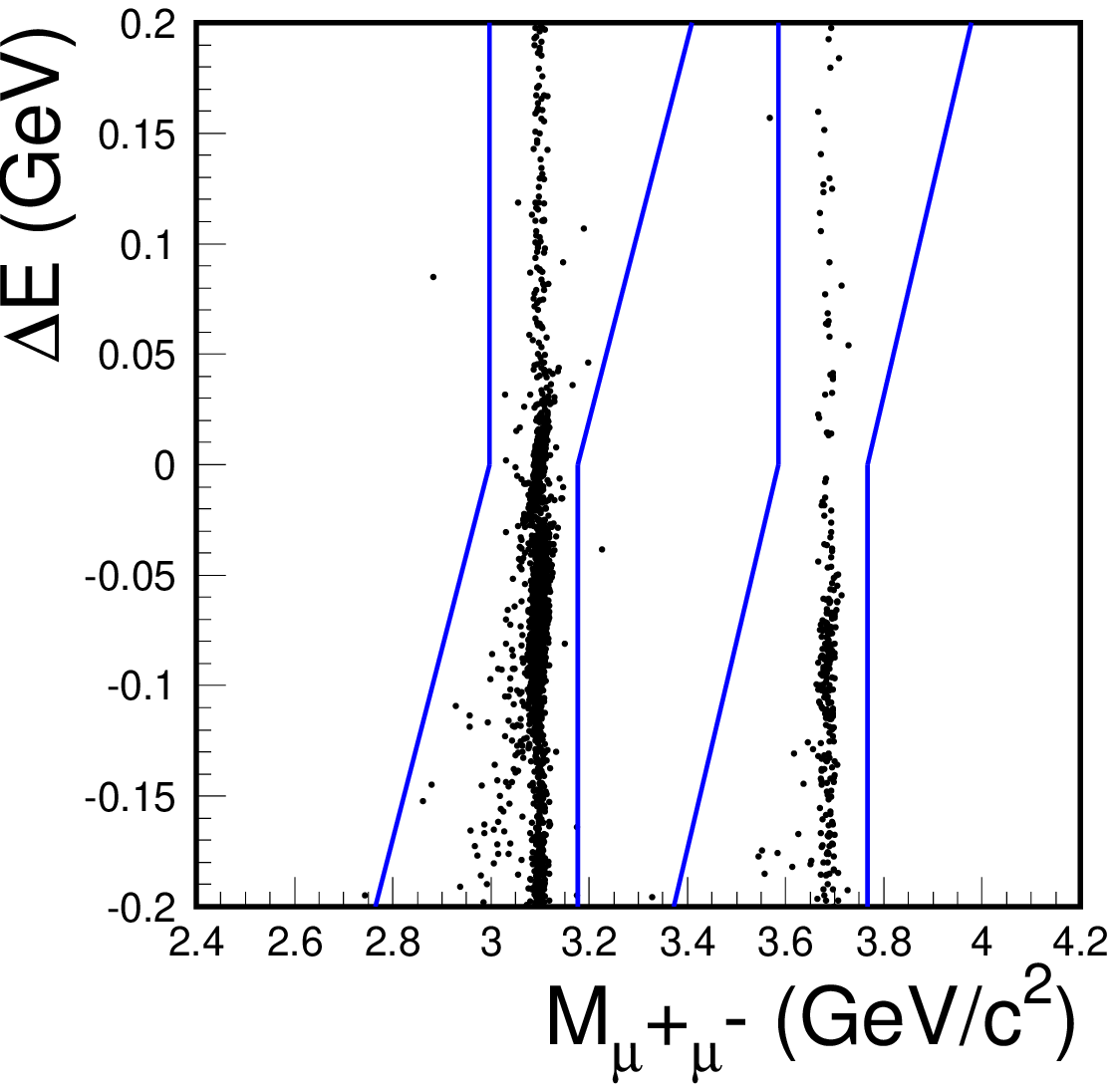}
%\hskip -0.3cm
\vskip -0.5cm
\includegraphics[width=5.5cm,height=5.5cm]{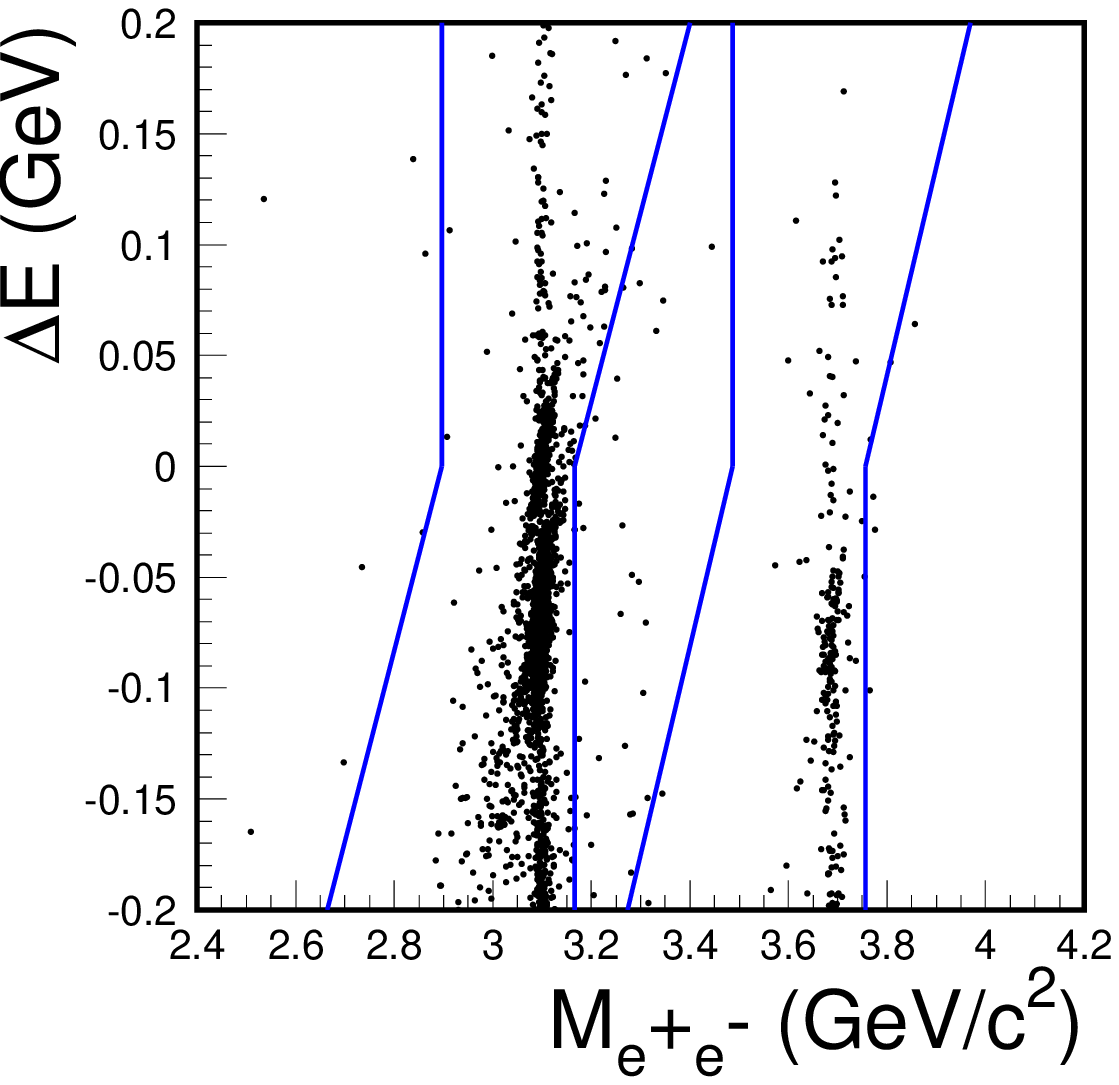}
\vskip -1.1cm
\end{center}
\caption{
$\Delta E$ vs. $M_{\ell^+\ell^-}$ distribution of 
$B \to J/\psi X$ and $\psi^\prime X$ MC events feeding into
$\pi^+ \mu^+ \mu^-$ (left) and $\pi^+ e^+ e^-$ (right) modes, respectively.}
\label{fig:devsmll}
\end{figure}

The decays $B \to J/\psi X$ and $\psi^\prime X$ are the dominant peaking backgrounds 
in the $\Delta E$--$M_{\mathrm{bc}}$ candidate region.
%The events in the following regions on $\Delta E$--$M_{\ell^+\ell^-}$ plane are rejected;
%the units are in GeV/$c^2$:
%(shown in Fig.~\ref{fig:devsmll})
Events in the regions 
\begin{eqnarray}
\nonumber -0.10 < &[M(\mu^+\mu^-)-m(J/\psi, \psi^\prime)]& < 0.08~, \\
\nonumber -0.20 < &[M(e^+e^-)-m(J/\psi, \psi^\prime)]& < 0.07~,
\end{eqnarray}
\begin{eqnarray}
\nonumber &&0.86\cdot[M(\mu^+\mu^-)-m(J/\psi, \psi^\prime)-0.08] < ~\\
\nonumber &&~~~~\Delta E/c^{2} < 0.86\cdot[M(\mu^+\mu^-)-m(J/\psi, \psi^\prime)+0.10]~,\\
%\nonumber ~{\rm and}~~\\
%\nonumber ~{\rm and}~~~~~~~~~ \\
\nonumber &&0.94\cdot[M(e^+e^-)-m(J/\psi, \psi^\prime)-0.07] <~\\
\nonumber &&~~~~\Delta E/c^{2} < 0.94\cdot[M(e^+e^-)-m(J/\psi, \psi^\prime)+0.20]~.
\end{eqnarray}
in the $\Delta E$--$M_{\ell^+\ell^-}$ plane are rejected 
(the limits on the regions are given in units of GeV/$c^2$, see Fig.~\ref{fig:devsmll}).
The decay $B^+ \to J/\psi(\psi^\prime) h^+$ ($h^+ = K^+,\pi^+$) 
%process has an extra contribution 
can also contribute 
to the $B^+\to \pi^+\mu^+\mu^-$ sample.
If a muon from $J/\psi(\psi^\prime)$ is misidentified as a pion and
%If a muon from $J/\psi(\psi^\prime)$ becomes a faked pion and
another non-muon track is at the same time misidentified as a muon, 
such a background event cannot be removed by the criteria described above.
%Thus, an additional veto region 
%$-0.10$ GeV/$c^2$ $< M(\mu \pi)-m(J/\psi, \psi^\prime) <$ $0.08$ GeV/$c^2$
%is introduced for the $B^+\to \pi^+\mu^+\mu^-$ decay.
Thus, in addition, events with
$-0.10$ GeV/$c^2$ $< M(\mu \pi)-m(J/\psi, \psi^\prime) <$ $0.08$ GeV/$c^2$
are removed from the $B^+\to \pi^+\mu^+\mu^-$ sample.
%The $B^0 \to D^-(\to \pi^0\pi^-) \pi^+$ has the chance to fake a $B^0\to \pi^0\mu^+\mu^-$ decay as well, if both pions are misidentified as muons.
%An event recording a $B^0 \to D^-(\to \pi^0\pi^-) \pi^+$ decay can look similar to
%$B^0\to \pi^0\mu^+\mu^-$ event, if both pions are misidentified as muons.
An event reconstructed as a $B^+ \to \overline{D}{}^0 (\to \pi^+\pi^-) \pi^+$ or 
$B^0 \to D^-(\to \pi^0\pi^-) \pi^+$ decay can contribute to the
$B\to \pi \mu^+\mu^-$ sample if both pions are misidentified as muons.
We reject these events by requiring $|M(\pi\mu)-m(D)| > $ 0.02 GeV/$c^2$.
The invariant mass of the electron pair is also required to be greater than 0.14 GeV/$c^2$ in order to remove 
background from photon conversions and from $\pi^0 \to \gamma e^+e^-$ decays. 
The background contributions from $B \to J/\psi(\psi^\prime) X$ and other $b\to c$ decays are estimated 
using large Monte Carlo (MC) samples. 
Background from charmless three-body $B$ decays, 
such as $B^+ \to \pi^+\pi^-\pi^+$ and $B^0 \to \pi^0\pi^-\pi^+$
is estimated using the measured data, 
taking into account the probabilities of the pions being misidentified as muons.
%incorporating with pion to muon fake rates.
The results are consistent with MC simulations.
The background yield from misidentified $B^+ \to K^+ \ell^+\ell^-$ decays is calculated from the
yield of a dedicated  $B^+ \to K^+ \ell^+\ell^-$ analysis with a similar analysis procedure,
which is then scaled by the kaon to pion misidentification rate.

If there are multiple candidates in an event, 
the candidate with the smallest vertex fit $\chi^{2}$ is chosen
for the $B^+ \to \pi^+ \ell^+ \ell^-$ modes,
and the one with the best $\mathcal{R}$ value is selected
for the $B^0 \to \pi^0 \ell^+ \ell^-$ modes.
%we select the candidate with the best $\mathcal{R}$ value.
The fraction of such events are about 11\%--13\% and 16\%--20\%
for $B^+ \to \pi^+ \ell^+ \ell^-$ and $B^0 \to \pi^0 \ell^+ \ell^-$, 
respectively, according to a MC study.
%and $16\sim20\%$ for $B^0 \to \pi^0 \ell^+ \ell^-$ from MC study.

We perform a simultaneous extended unbinned maximum likelihood fit
to $\Delta E$ and $M_{\rm bc}$
%The signal yields are extracted by maximizing 
with the following likelihood function:
\begin{eqnarray}
\nonumber
\mathcal{L} & = & {e^{-(N_s+N_{q\overline{q}}+N_{c\overline{c} X}+N_{K\ell\ell}+N_{h \pi \pi})}\over N!} \times \\
\nonumber
&&\prod_{i=1}^{N}~[
%N_s P_s(M_{{\rm bc}_i},\Delta{E}_i)+ \\
%\nonumber
%&& N_{q\overline{q}} P_{q\overline{q}} (M_{{\rm bc}_i},\Delta{E}_i) + \\
%\nonumber
%&& N_{B\overline{B}} P_{B\overline{B}} (M_{{\rm bc}_i},\Delta{E}_i) + \\
%\nonumber
%&& N_{c\overline{c} X} P_{c\overline{c} X} (M_{{\rm bc}_i},\Delta{E}_i) + \\
%\nonumber
%&& N_{K\ell\ell} P_{K\ell\ell} (M_{{\rm bc}_i},\Delta{E}_i) +  \\
%&& N_{h \pi \pi} P_{h \pi \pi} (M_{{\rm bc}_i},\Delta{E}_i)]~.~~~~~~
N_s P_s^{i}+ N_{q\overline{q}} P_{q\overline{q}}^{i} + N_{B\overline{B}} P_{B\overline{B}}^{i} + \\
\nonumber
&& N_{c\overline{c} X} P_{c\overline{c} X}^{i} + N_{K\ell\ell} P_{K\ell\ell}^{i} + N_{h \pi \pi} P_{h \pi \pi}^{i}]~.~~~~~~
\end{eqnarray}
%where $P_s$, $P_{q\overline{q}}$, $P_{B\overline{B}}$, $P_{c\overline{c} X}$, 
%$P_{K\ell\ell}$, and $P_{h \pi \pi}$ denote the probability density
%functions (PDFs) 
where $N$ denotes the number of observed events in the candidate region, 
and $N_s$ ($P_s^{i}$), $N_{q\overline{q}}$ ($P_{q\overline{q}}^{i}$), 
$N_{B\overline{B}}$ ($P_{B\overline{B}}^{i}$), $N_{c\overline{c} X}$ ($P_{c\overline{c} X}^{i}$), 
$N_{K\ell\ell}$ ($P_{K\ell\ell}^{i}$), and $N_{h \pi \pi}$ ($P_{h \pi \pi}^{i}$) 
denote the event yields (the probability density functions, PDFs, for the i-th event)  
for signal, continuum, $b \to c$ decays, $B \to J/\psi(\psi^\prime) X$, 
background contributions from $B \to K \ell^+\ell^-$ and $B \to h \pi^+\pi^-$ decays.
Since contributions from $B \to J/\psi(\psi^\prime) X$ and $B \to K \ell^+\ell^-$ processes are 
found to be negligible for $B^0 \to \pi^0 \ell^+\ell^-$ decays, the associated PDFs are excluded from the fits.
The signal PDFs for $B^+ \to \pi^+ \ell^+\ell^-$ 
are assumed to be Gaussian in both $\Delta E$ and $M_{\rm bc}$.
The means and widths are
verified using $B \to J/\psi K$ decays. 
Additional Gaussian functions are used to model the tails in the $\Delta E$ distributions.
For the $B^0 \to \pi^0 \ell^+ \ell^-$ modes, the signal PDF is modeled by a two-dimensional 
smoothed histogram function, including the dependence on $M(\ell^+ \ell^-)$.
The continuum PDFs for $M_{\rm bc}$ and $\Delta E$ are represented by
an empirical background function introduced by ARGUS \cite{ref:argus}
and a second-order polynomial, respectively.
The PDF for $B \to J/\psi(\psi^\prime) X$ decays consists of a peaking part and a non-peaking part. 
The peaking PDF is modeled by Gaussian functions in $\Delta E$ and $M_{\rm bc}$.
The non-peaking part, as well as the PDF for other $b \to c$ decays, 
is described by the same PDF formulae used for continuum events, 
but with different parameter values. 
The distributions of $\Delta E$ and $M_{\rm bc}$ for the $B \to K \ell^+\ell^-$ and $B \to h \pi^+\pi^-$ contributions to
the $B^+ \to \pi^+ \ell^+\ell^-$ modes are found to be peaking and modeled by Gaussian functions. 
Due to a much larger tail on $\Delta E$ for the $B^0 \to \pi^0 \ell^+\ell^-$ decays, 
the $B \to h \pi^+\pi^-$ contributions to $B^0 \to \pi^0 \ell^+\ell^-$ are modeled by smoothed histogram functions.

\begin{figure*}[t!]
%\begin{figure*}[hb]
\begin{center}
\includegraphics[width=3.8cm,height=7.2cm]{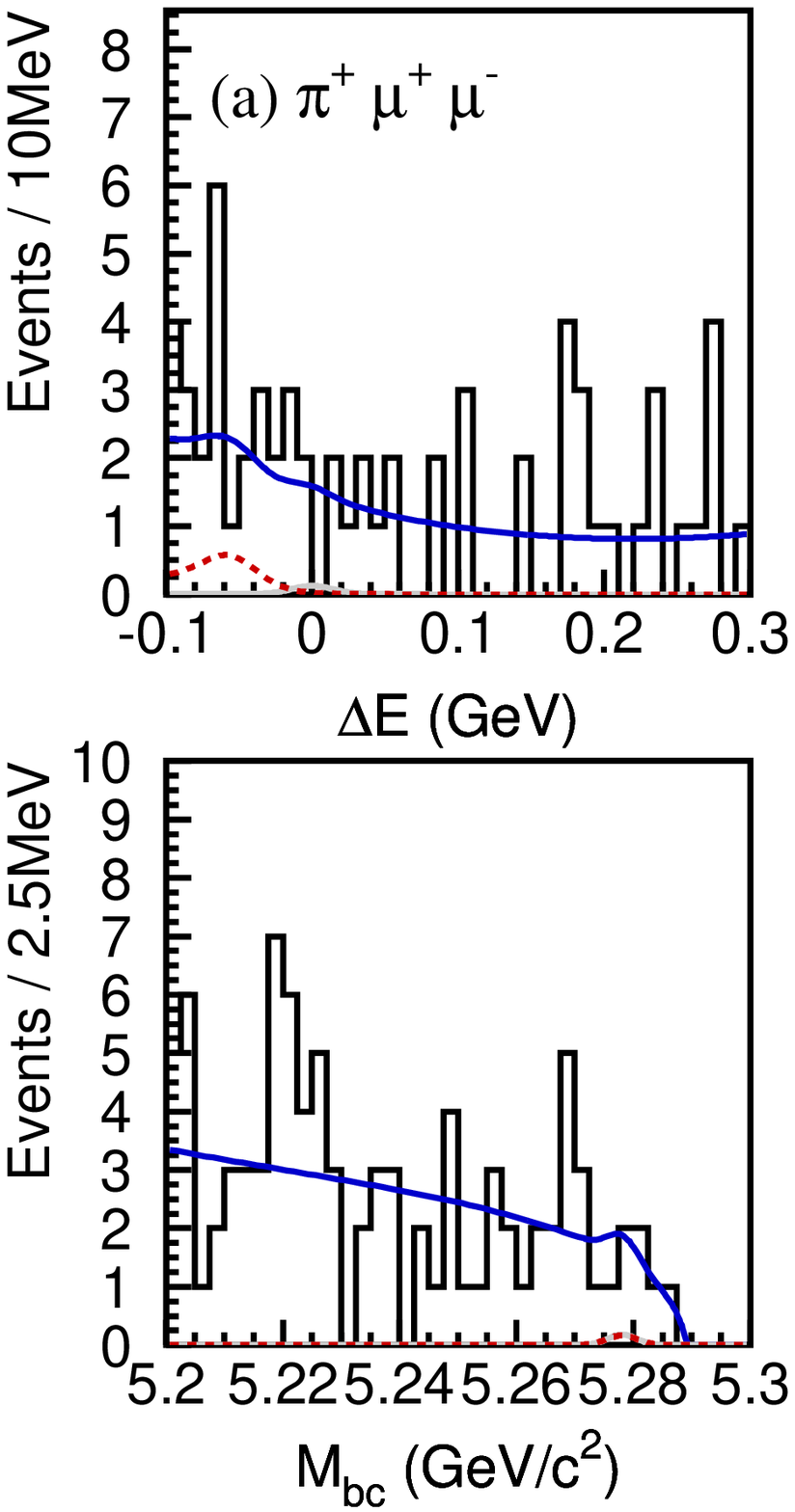}
\includegraphics[width=3.8cm,height=7.2cm]{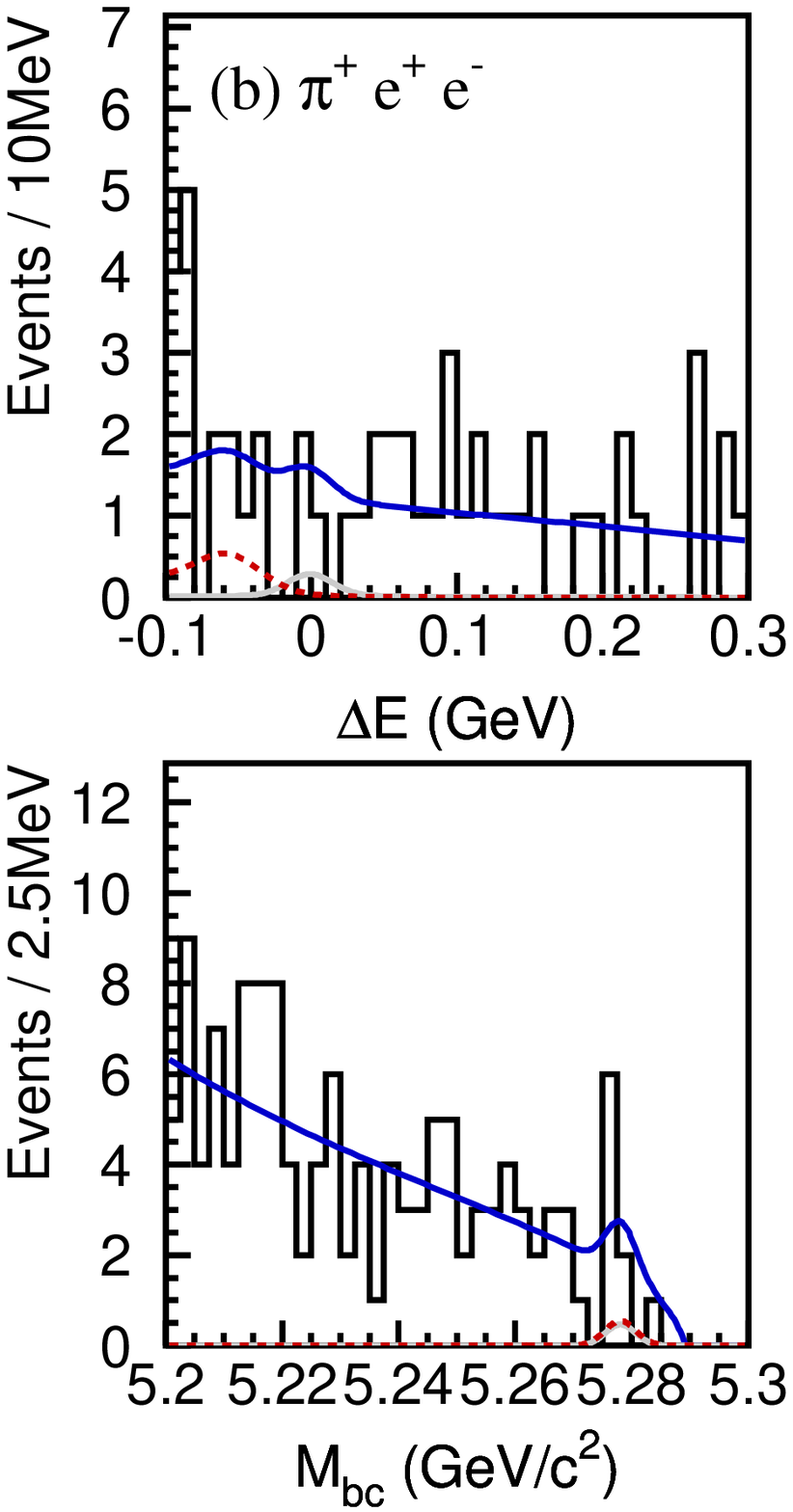}
\includegraphics[width=3.8cm,height=7.2cm]{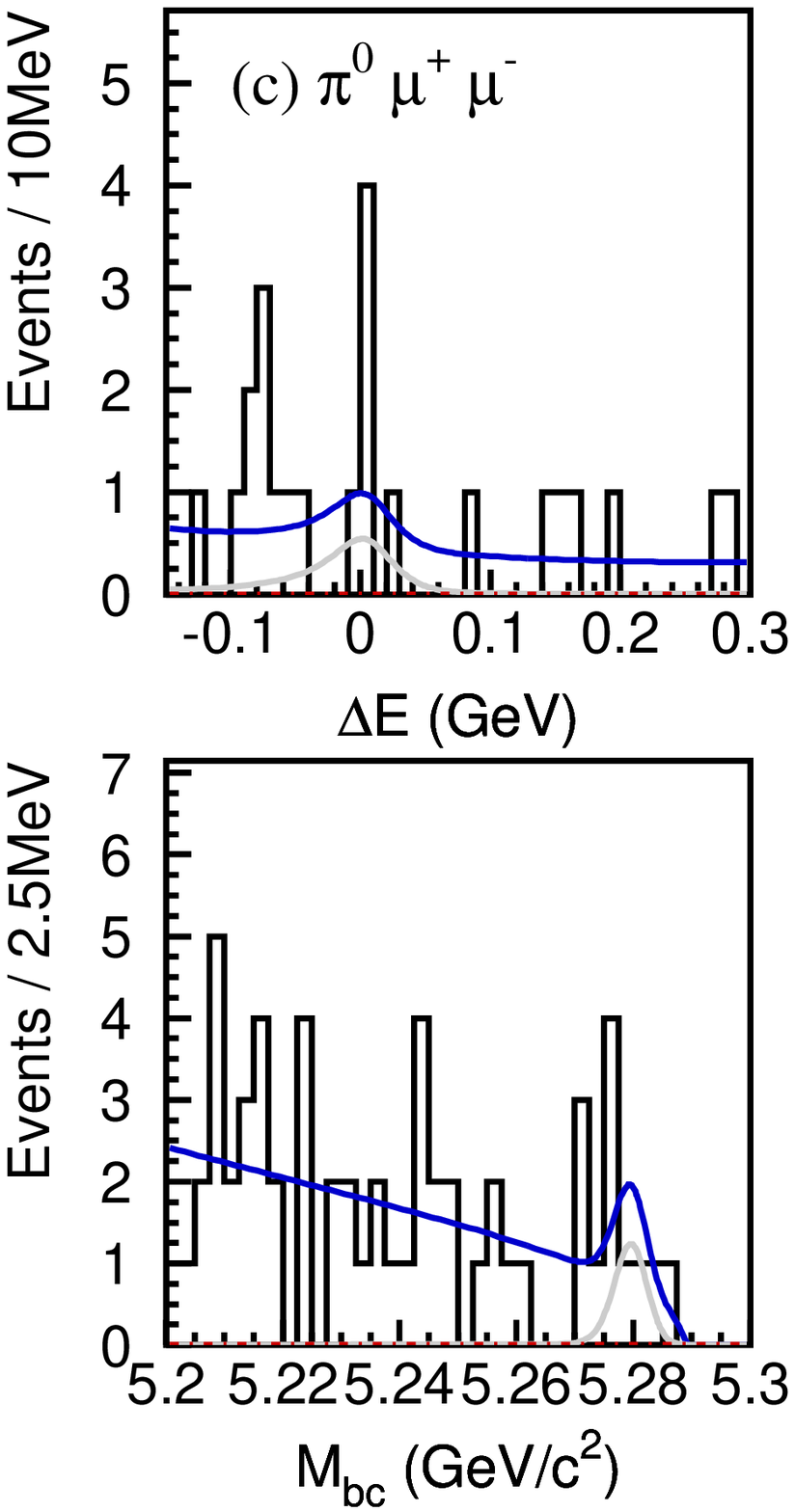}
\includegraphics[width=3.8cm,height=7.2cm]{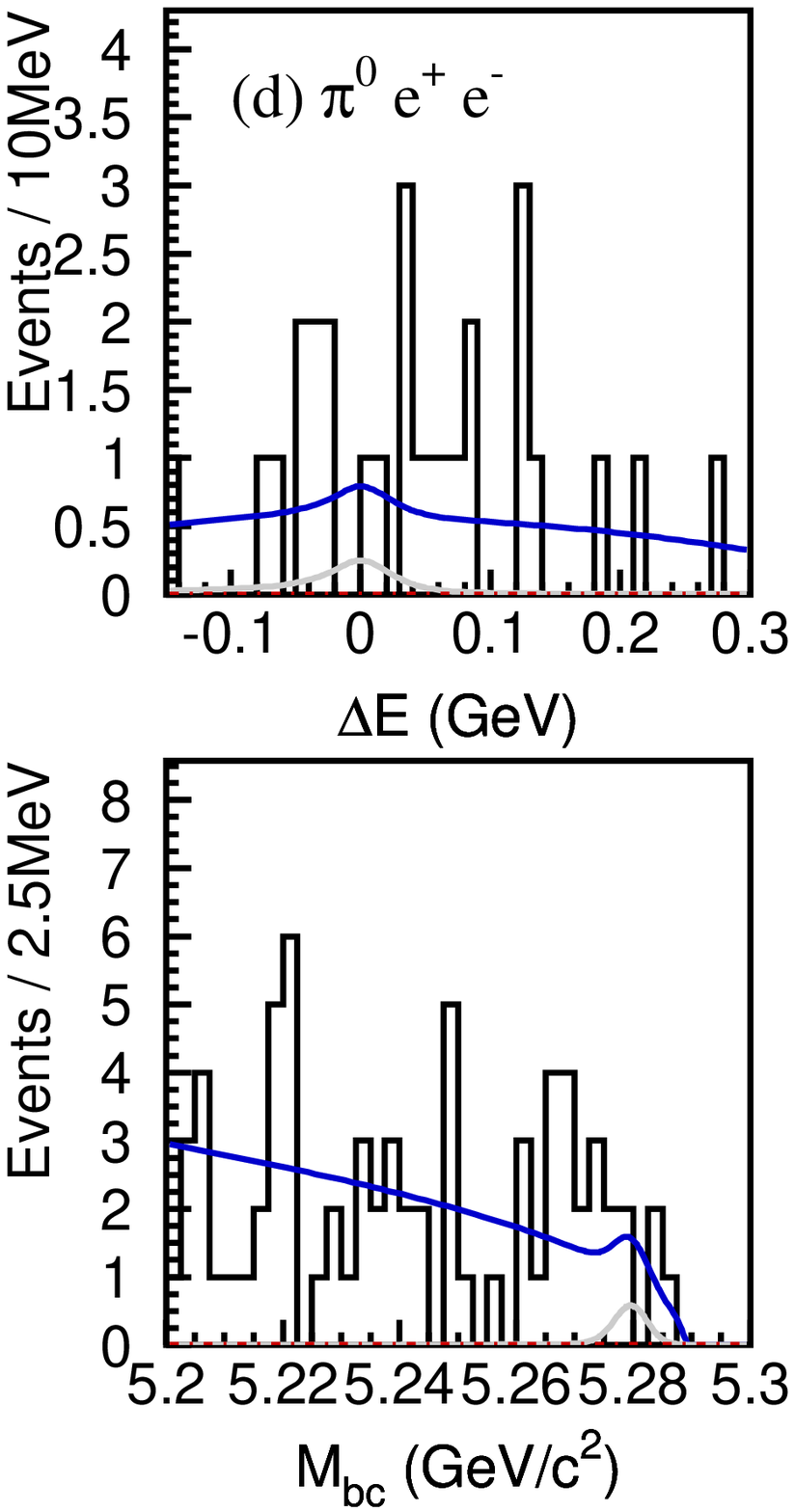}
\end{center}
\caption{
Distributions of $\Delta E$ ($M_{\rm bc}$) with fit results superimposed for
the events in the $M_{\rm bc}$ ($\Delta E$) signal region.
%the events in the region of 5.27~GeV/$c^2 < M_{\rm bc} <$ 5.29~GeV/$c^2$
%(-0.035, -0.055, -0.08, and -0.1 GeV $< \Delta E <$ 0.035 GeV for the 
%$\pi^+\mu^+\mu^-$, $\pi^+e^+e^-$, $\pi^0\mu^+\mu^-$, and $\pi^0e^+e^-$ modes, respectively)
The solid curves represent the fit results, 
%with the signal ($K^+ \ell^+\ell^-$ background) peaks represented by the solid (dashed) peaks.
while the solid (dashed) peaks represent the signal ($K^+ \ell^+\ell^-$ background) component.
%and the background components from $B^+ \to K^+ \ell^+\ell^-$ are shown by the dashed curves.
}
\label{fig:projections}
\end{figure*}

Yields for signal and continuum, and the continuum PDF parameters
are allowed to float in the fit while 
the yields and parameters for other components are fixed.
The observed signal yields and branching fractions (${\cal B}$) are summarized in Table~\ref{tab:results}.
The significance is defined as $\sqrt{-\rm{2ln} \left( \mathcal{L}_{0} / \mathcal{L}_{\rm max} \right)}$,
where $\mathcal{L}_{0}$ is the likelihood with signal yield constrained to be zero, 
and $\mathcal{L}_{\rm max}$ is the maximum likelihood.
%The reconstruction efficiencies are estimated by MC simulations according to Ref.~\ref{}.
The distributions of $\Delta E$ and $M_{\rm bc}$ with fit results superimposed 
are shown in Fig.~\ref{fig:projections}.

\begin{table*}[t!]
%\begin{table*}[b!]
\caption{A summary of the signal yields ($N_s$), reconstruction efficiencies ($\epsilon$), 
statistical significance ($\Sigma$), branching fractions ($\mathcal{B}$), 
and the corresponding upper limits (U.L.) at the 90\% confidence level.
%\textcolor{blue}{
The errors associated with branching fractions are 
sequentially statistical errors, 
systematic errors proportional to the branching fractions (given in Table II), 
and additive systematic errors related to yield extraction.%}
}
\label{tab:results}
\begin{center}
\begin{tabular}{lccccc}
\hline
Mode~ & $N_s$ & $\epsilon$ (\%) & $\Sigma$ & $\mathcal{B}$ ($10^{-8}$) & U.L. ($10^{-8}$) \\ 
\hline
$B^+ \to \pi^+\mu^+\mu^-$   & $0.5^{+2.8}_{-1.9}$~ & 13.1~ & 0.2$\sigma$~ &
 $0.6^{+3.2}_{-2.2}\pm$0.0$\pm$0.7~ & 6.9~ \\
$B^+ \to \pi^+e^+e^-$       & $1.4^{+3.2}_{-2.3}$~ & 13.8~ & 0.6$\sigma$~ &
 $1.5^{+3.5}_{-2.5}\pm$0.1$\pm$0.8~ & 8.0~ \\
$B^+ \to \pi^+\ell^+\ell^-$ & -                  ~ & -   ~ & 0.6$\sigma$~ & 
 $1.0^{+2.3}_{-1.8}\pm$0.1$\pm$0.4~ & 4.9~ \\
\hline
$B^0 \to \pi^0\mu^+\mu^-$   & $5.1^{+4.2}_{-3.3}$~ &  9.6~ & 1.8$\sigma$~ &
 $8.1^{+6.7}_{-5.2}\pm$0.8$\pm$1.0~ & 18.4~ \\
$B^0 \to \pi^0e^+e^-$       & $2.7^{+5.2}_{-4.0}$~ &  7.4~ & 0.7$\sigma$~ &
 $5.5^{+10.7}_{-8.3}\pm$0.5$\pm$2.0~ & 22.7~ \\
$B^0 \to \pi^0\ell^+\ell^-$ & -                    &  -    & 2.0$\sigma$~ &
 $7.4^{+5.4}_{-4.5}\pm$0.7$\pm$0.8~ & 15.4~ \\
\hline
$B   \to \pi\ell^+\ell^-$   & -                    &  -    & 1.2$\sigma$~ &
 $2.4^{+2.5}_{-2.1}\pm$0.2$\pm$0.2~ & 6.2~ \\
\hline
\end{tabular}
\end{center}
\end{table*} 

Systematic uncertainties are summarized in Table~\ref{tab:systematics}.
The dominant systematic uncertainties for the $B^+ \to \pi^+\ell^+\ell^-$ modes stem from 
electron (4.4\%) and muon (5.5\%) identification efficiencies, 
uncertainties of MC decay models (2.6\%--3.8\%), and tracking efficiencies (3\%). 
%The MC modeling uncertainties are studied by comparing the $B^+ \to K^+\ell^+\ell^-$ yields 
%as a function of the dilepton invariant mass between MC and data.
The signal MC samples are generated based on the decay model derived from~\cite{Ali:2002jg}, 
and the uncertainties are evaluated by comparing the MC and data yields for
$B^+ \to K^+\ell^+\ell^-$ as a function of dilepton invariant mass.
Other uncertainties such as background suppression, 
which is studied using $B \to J/\psi K$ decays, 
and PDF modeling are all small.
For the $B^0 \to \pi^0\ell^+\ell^-$ modes, the dominant uncertainties are from lepton identifications, 
PDF modeling (5.3\%), $\pi^0$ reconstruction efficiency (4\%), and MC models.
Other systematic uncertainties are found to be small.
%\textcolor{blue}{
There are also some small additive uncertainties that are not proportional to 
the signal branching fractions. 
An example is the uncertainty in the background level, 
which is estimated to be about 0.3-0.8 events.
In addition, a fitting bias is found in the MC simulation 
when signal yields are small (fewer than 2 events). 
Thus, we quote an additional uncertainty of 0.5 events for each mode.
%such as background uncertainty, which is estimated to be about 0.3--0.8 events.
%Besides, a fitting bias is found in MC study when signal yields are few (less than 2 events).
%}

\begin{table*}[t!]
%\begin{table*}[b!]
\caption{Summary of the contributions of systematic uncertainty proportional to the branching fractions (in \%).}
\label{tab:systematics}
\begin{center}
\begin{tabular}{lcccc}
\hline
Decay & $B^+\to\pi^+\mu^+\mu^-$ & $B^+\to\pi^+e^+e^-$ & $B^0\to\pi^0\mu^+\mu^-$ & $B^0\to\pi^0e^+e^-$ \\
\hline
Tracking efficiency                & 3.0 & 3.0 & 2.0 & 2.0 \\
$e$/$\mu$-identification           & 5.5 & 4.4 & 5.5 & 4.4 \\
$\pi^+$-id./$\pi^0$ reconstruction & 1.0 & 1.0 & 4.0 & 4.0 \\
Background suppression             & 1.8 & 1.8 & 1.8 & 2.4 \\
PDF Modeling                       & 2.2 & 2.2 & 5.3 & 5.3 \\
MC decay model                     & 3.8 & 2.6 & 3.8 & 2.6 \\
MC statistics                      & 0.8 & 0.8 & 1.1 & 1.1 \\
$N(B\overline{B})$ pairs           & 1.3 & 1.3 & 1.3 & 1.3 \\
\hline
Total                              & 8.1 & 6.8 & 10.0& 9.1 \\
\hline
\end{tabular}
\end{center}
\end{table*}

As no statistically significant excess of signal events is observed,
we calculate upper limits of the branching fractions 
by integrating the likelihood function up to the 90\% confidence level.
We find,
\begin{eqnarray}
\nonumber
\mathcal{B}(B^+ \to \pi^+\mu^+\mu^-) &<& 6.9 \times 10^{-8}~,\\
\nonumber
\mathcal{B}(B^+ \to \pi^+e^+e^-)     &<& 8.0 \times 10^{-8}~,\\
\nonumber
\mathcal{B}(B^0 \to \pi^0\mu^+\mu^-) &<& 18.4 \times 10^{-8}~,\\
\mathcal{B}(B^0 \to \pi^0e^+e^-)     &<& 22.7 \times 10^{-8}~.
\end{eqnarray}
Systematic uncertainties are included by convoluting the likelihood function with a Gaussian function.
%The effect of systematic uncertainties is included 
%by smearing the likelihood function with a gaussian in the calculation.
%to include the effect of systematic uncertainties in the calculation.
%which width is the total systematic uncertainty.
The lepton-flavor combined limits are given by 
\begin{eqnarray}
\nonumber
\mathcal{B}(B^+ \to \pi^+\ell^+\ell^-) &<& 4.9 \times 10^{-8}~,\\
\mathcal{B}(B^0 \to \pi^0\ell^+\ell^-) &<& 15.4 \times 10^{-8}~.
\end{eqnarray}
%Assuming isospin symmetry, the average branching fraction and limit for $B \to \pi\ell^+\ell^-$ 
%are estimated to be
	
Assuming isospin symmetry, the average limit for $B \to \pi\ell^+\ell^-$ is
\begin{eqnarray}
\nonumber
%&&\mathcal{B}(B^+ \to \pi^+\ell^+\ell^-) \\
%\nonumber
%&&= 2 \times {\tau_{B^+} \over \tau_{B^0}}\mathcal{B}(B^0 \to \pi^0\ell^+\ell^-) \\
%\nonumber
%&&= (0.24^{+0.25}_{-0.21}\pm 0.03) \times 10^{-7} \\
%&&< 0.62 \times 10^{-7}~. 
\mathcal{B}(B^+ \to \pi^+\ell^+\ell^-) &<& 6.2 \times 10^{-8}~.
\end{eqnarray}

In conclusion, 
%we present a search for $B \to \pi\ell^+\ell^-$ decays using a sample of 
%657 million $B\overline{B}$ pairs collected at the $\Upsilon(4S)$ resonance. 
%The average $B \to \pi\ell^+\ell^-$ branching fraction is 
%consistent with the prediction of the SM.
%The combined statistical significance is only 1.2 standard deviations.
the upper limit on the isospin-averaged branching fraction is 
about twice the SM expectation,
and two thirds of the BaBar measurement~\cite{ref:pill_exp}.
A much larger data set, such as will be available from the proposed super $B$ factory~\cite{ref:superB}, 
is needed to observe this decay mode
if the branching fraction is at the level
%make an actual observation of the decay mode as 
predicted by the SM.

% Please paste this acknowledgement into your latex file. 
% updated 2/5/08     Super SINET -> SINET3
% corrected 10/26/07 China missed "Natural" is recovered (Long only)
% updated 9/18/07    China, KIP of CAS removed, new cont. no.
% updated 1/23/07    add RFAE and change MIST -> MES for Russia
% updated 11/22/06   Chinese Academy of Sciencies -> Sciences
% updated 6/23/06
% updated 2/21/06, 1/28/06, 12/25/05
% short version reduced 8/11/05
% updated 7/17/05, 2/17/05
%***** Acknowledgments *****
%----------- Long version, for most papers ----------- 
We thank the KEKB group for the excellent operation of the
accelerator, the KEK cryogenics group for the efficient
operation of the solenoid, and the KEK computer group and
the National Institute of Informatics for valuable computing
and SINET3 network support. We acknowledge support from
the Ministry of Education, Culture, Sports, Science, and
Technology of Japan and the Japan Society for the Promotion
of Science; the Australian Research Council and the
Australian Department of Education, Science and Training;
the National Natural Science Foundation of China under
contract No.~10575109 and 10775142; the Department of
Science and Technology of India; 
the BK21 program of the Ministry of Education of Korea, 
the CHEP SRC program and Basic Research program 
(grant No.~R01-2005-000-10089-0) of the Korea Science and
Engineering Foundation, and the Pure Basic Research Group 
program of the Korea Research Foundation; 
the Polish State Committee for Scientific Research; 
%-> remove for now: under contract No.~2P03B 01324; 
the Ministry of Education and Science of the Russian
Federation and the Russian Federal Agency for Atomic Energy;
the Slovenian Research Agency;  the Swiss
National Science Foundation; the National Science Council
and the Ministry of Education of Taiwan; and the U.S.\
Department of Energy.
% -------------------------------------------------

\end{document}